\documentclass[preprint,12pt]{elsarticle}
\usepackage{amssymb,amsmath}
\journal{Journal of Theoretical Biology}

\begin{document}

\begin{frontmatter}

\title{Combination with anti-tit-for-tat remedies problems of tit-for-tat}
\author[sdy]{Su Do Yi}
\author[skb]{Seung Ki Baek\corref{cor1}}
\ead{seungki@pknu.ac.kr}
\author[jkc]{Jung-Kyoo Choi\corref{cor2}}
\ead{jkchoi@knu.ac.kr}
\cortext[cor1]{Principal corresponding author}
\cortext[cor2]{Corresponding author}

\address[sdy]{Department of Physics and Astronomy, Seoul National University,
Seoul 08826, Korea}
\address[skb]{Department of Physics, Pukyong National University, Busan 48513,
Korea}
\address[jkc]{School of Economics and Trade, Kyungpook National University,
Daegu 41566, Korea}

\begin{abstract}
One of the most important questions in game theory concerns how mutual
cooperation can be achieved and maintained in a social dilemma.  In Axelrod's
tournaments of the iterated prisoner's dilemma, Tit-for-Tat (TFT) demonstrated
the role of reciprocity in the emergence of cooperation. However, the stability
of TFT does not hold in the presence of implementation error, and a TFT
population is prone to neutral drift to unconditional cooperation, which
eventually invites defectors.  We argue that a combination of TFT and anti-TFT
(ATFT) overcomes these difficulties in a noisy environment, provided that ATFT
is defined as choosing the opposite to the opponent's last move.  According to
this TFT-ATFT strategy, a player normally uses TFT; turns to ATFT upon
recognizing his or her own error; returns to TFT either when mutual cooperation
is recovered or when the opponent unilaterally defects twice in a
row. The proposed strategy provides simple and deterministic behavioral
rules for correcting implementation error in a way that cannot be exploited by
the opponent, and suppresses the neutral drift to unconditional cooperation.
\end{abstract}


\end{frontmatter}

\section{Introduction}

Game theory has provided useful frameworks to understand conflict and
cooperation in the absence of a central authority.
In many studies, reciprocity is a particularly important idea to explain
cooperation between rational players.
In the iterated version of the Prisoner's Dilemma (IPD) game,
the repetition makes it profitable for each player to cooperate because
one has to take the other's reaction into
account~\cite{Sethi,Nowak,Szabo,Sigmund,Perc,Compare}.
As long as both are sufficiently patient and free from error,
therefore, the players cooperate from the first encounter
and this continues forever.
The notion of reciprocity has been popularized by the success of
Tit-for-Tat (TFT) in Axelrod's tournaments~\cite{Axelrod}.
TFT cooperates at the first encounter with the co-player, and replicates the
co-player's previous move in the subsequent rounds. Axelrod has argued that this
simple strategy is nice, retaliating, forgiving, and non-envious.
However, the situation becomes much more complicated in a noisy environment,
where it is possible to make an implementation error: For example,
there can be an inefficient alternation of cooperation and defection between two
TFT players when one of them makes a mistake~\cite{Molander,Boyd}.
Tit-for-Two-Tats (TF2T) is more tolerant, because it retaliates when the
co-player defects twice in a row. Although TF2T was proposed as a remedy
to avoid the inefficiency of TFT, its predictability opens another possibility
of being exploited.
Generous TFT avoids this dilemma between generosity and exploitability
by introducing randomness in
forgiving the co-player's defection~\cite{Nowak92,Imhof05,Imhof07,Imhof10}.
However, its unpredictability might be a double-edged sword when it comes to
public policy making, in which such random factors are not popular
ideas~\cite{Dror}.
Moreover, Generous TFT is overwhelmed by a more generous strategy called
Win-Stay-Lose-Shift (WSLS), also known as
Pavlov~\cite{Kraines}, in the evolution of cooperation~\cite{Nowak93}.
WSLS is one of aspiration-based strategies~\cite{Posch,Liu},
which means that it attempts changes when its payoff becomes less than a
certain level. For example, when WSLS is in mutual defection with its
co-player, it receives a lower payoff than the aspiration level and thus
cooperates next time.
The success of WSLS originates from
the ability to correct error quickly when played against another WSLS player.
The price is that WSLS easily falls into a victim of an unconditional
defector (AllD). All these difficulties are rooted in the error vulnerability
of TFT. Worse is that this is not the only weakness:
A TFT population is invaded via neutral drift by
unconditional cooperators (AllC), which, in turn, opens the back door to
AllD~\cite{Imhof05,Imhof07,Imhof10,Young}. It would thus be desirable if
some modification overcame these shortcomings while preserving the strengths of
TFT. This work will show that such a combination is actually possible.

A method of searching for successful cooperating strategies is to perform an
evolutionary experiment {\it in silico}~\cite{Nowak93,Axelrod87,Lindgren}.
The purpose of such an experiment is to simulate natural selection.
The experimenter does not have direct control over selection,
but only determines evolutionary dynamics and an accessible set of strategies.
If this can be called a bottom-up approach, it is
also possible to think of a top-down approach;
we can impose certain criteria that a successful strategy is
expected to
satisfy and sort out strategies that meet the criteria.
This work follows the latter approach.
One of its advantages is that it does not need pairwise comparisons between
all the strategies under consideration so that the computational cost scales
only linearly with the number of strategies.
We do not mean that this approach will find an optimal strategy such that
guarantees maximal payoffs. Nor do we mean that it will be equivalent to
the result of evolutionary experiments or actual human behavior.
Rather, our purpose is to \emph{design} a workable solution
to avoid a series of wasting retaliation from a mistaken move,
when players are bound to interact repeatedly.

Our finding is that the top-down approach singles out a strategy that combines
TFT and anti-TFT (ATFT) in a way that it stabilizes mutual cooperation in the
presence of implementation error.
Note that ATFT does the opposite to the co-player's last move.
According to this proposed strategy, our focal player Alice normally uses TFT
but switches to ATFT when she recognizes her own error.
Alice returns back to TFT either when mutual
cooperation is recovered or when Bob unilaterally defects twice in
a row.
The strategy says that Alice is responsible only for her own error without any
need to judge Bob's behavior. Our result provides a simple deterministic
error-correcting rule, which is secured against repeated exploitation at the
same time.

Before proceeding to the next section, let us assume that the players in this
work perceive their respective payoffs with no uncertainty as defined by the
game, although they do make implementation errors.
In other words, their history of moves is given as public information,
whose accessibility is limited only by their memory spans. Moreover, 
we set aside other public information such as social standing,
which is required in Contrite
TFT~\cite{Sugden,Wu,Boerlijst,Hilbe09} as well as in some form of indirect
reciprocity~\cite{Nowak98,Panchanathan,Ohtsuki04,Ohtsuki05,Brandt,Uchida,Olejarz}.
After deriving our main results under these assumptions, we will revisit the
problem of perception error.

\section{Method and Result}
\label{sec:method}

The payoff matrix $M$ of the PD game can be written as follows:
\begin{equation}
\left(
\begin{array}{c|cc}
  & C & D\\\hline
C & M_{CC} & M_{CD}\\
D & M_{DC} & M_{DD}
\end{array}
\right),
\label{eq:payoff}
\end{equation}
where $C$ and $D$ denote each player's possible moves, cooperation and
defection, respectively.
The matrix elements satisfy two inequalities: First, $M_{DC} > M_{CC}
> M_{DD} > M_{CD}$ for mutual defection to be a Nash equilibrium. Second,
$M_{CC} > (M_{CD} + M_{DC}) / 2$ for mutual cooperation to be Pareto optimal.

In the IPD, we define a state as a history of moves chosen by two players
Alice and Bob for the past $m$ time steps.
The number of states is $n=2^{2m}$, because each of the two players has
$m$ binary choices between $C$ and $D$.
Choosing $m=2$, for example,
we work with $n=16$ states, each of which at time $t$ can be written as
$S_t = (A_{t-2} A_{t-1}, B_{t-2} B_{t-1}) =$
(Alice's moves at $t-2$ and $t-1$, Bob's moves at $t-2$ and $t-1$).
Alice's strategy is
represented as a mapping from the state $S_t$ to her move $A_t \in \{C,D\}$ at
an arbitrary time step $t$. The total number of Alice's
possible strategies is thus $N=2^n = 2^{2^{2m}}$.
Here, we do not specify initial moves for $t \le m$ in
defining a strategy, because they are irrelevant when we consider long-term
averages in the presence of error, as will be explained below.

\begin{figure}
\begin{center}
\centerline{\includegraphics[width=0.4\textwidth]{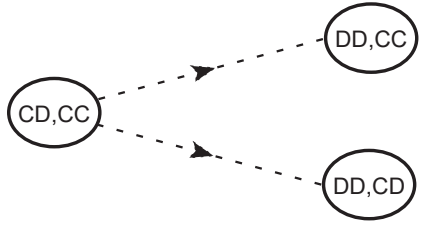}}
\caption{Possible transitions from state
$S_t = (A_{t-2} A_{t-1}, B_{t-2} B_{t-1}) = (CD, CC)$ to $S_{t+1} = (A_{t-1}
A_{t}, B_{t-1} B_{t})$, which can be either $(DD,CC)$ or $(DD,CD)$ when Alice's
strategy prescribes $A_t = D$ for that $S_t$.
}\label{fig:transit}
\end{center}
\end{figure}

It is instructive to note that the state at time $t+1$ is written as
$S_{t+1} = (A_{t-1} A_{t}, B_{t-1} B_{t})$, which
always shares $A_{t-1}$ and $B_{t-1}$ with $S_t = (A_{t-2} A_{t-1}, B_{t-2}
B_{t-1})$. In
addition, $A_t$ is determined from $S_t$ by our focal player Alice's strategy.
For this reason, when Alice's strategy and state $S_t$ are given, the only
unknown part is $B_t \in \{C,D\}$.  We impose no restriction on Bob's
strategy, so each state can generally be followed by one of two different
states at the next time step. For example, suppose that
Alice finds herself in $S_t = (A_{t-2} A_{t-1}, B_{t-2} B_{t-1}) = (CD, CC)$ at
time $t$ and her strategy prescribes $A_t = D$ for this $S_t$. Then, we
conclude that her next state $S_{t+1} = (A_{t-1} A_{t}, B_{t-1} B_{t})$ must be
either $(DD, CC)$ or $(DD, CD)$. Graphically, this idea can be represented as
depicted in Fig.~\ref{fig:transit}.

Suppose a sequence of moves in the IPD between two strategies $i$
and $j$. Let $F_{ij}^{(t)} \in \{M_{CC}, M_{CD}, M_{DC}, M_{DD} \}$ denote the
payoff of strategy $i$ against $j$ at time step $t$. We are interested in its
long-term average, i.e.,
\begin{equation}
\overline{F_{ij}}(T) = \frac{1}{T} \sum_{t=1}^T F_{ij}^{(t)}
\end{equation}
with $T \gg 1$.
If both $i$ and $j$ have finite memories, the calculation of $\overline{F_{ij}}
(T \rightarrow \infty)$
can be simplified as follows:
Any pair of strategies with finite memories constitute
a Markov process described by a stochastic matrix. When error is absent,
the moves of two deterministic strategies eventually form a loop.
By a loop, we
mean a sequence of consecutive states $S_t \rightarrow S_{t+1}
\rightarrow \ldots \rightarrow S_{t+\nu}$ allowed by the strategies, in which
$S_{t+\nu} = S_t$ with a positive integer $\nu$.
The integer $\nu$ is called the length of the loop.
If the number of possible loops is greater than one, which loop the players will
be trapped in will be determined by an initial state, which corresponds to their
first moves. However, when implementation error occurs with probability $e \ll
1$ so that intended cooperation or defection can fail,
they can escape from the loop with a time scale of $O(1/e)$.
The probabilities assigned to states thus converge to a unique
stationary distribution in the long run, regardless of the initial
state~\cite{Nowak90,Nowak95}.
For example, when AllC meets TFT, the interaction can be described by
the following $n \times n$ stochastic matrix $U$ with $n=2^{2m} = 4$:
\begin{eqnarray}
&\begin{array}{cccc}
j=(C,C) & ~~(C,D) & ~~(D,C) & ~~(D,D)~~~~~~
\end{array}\nonumber\\
i=
\begin{array}{c}
(C,C)\\
(C,D)\\
(D,C)\\
(D,D)
\end{array}
&\left(
\begin{array}{cccc}
(1-e)^2 & (1-e)^2 & (1-e)e & (1-e)e\\
(1-e)e & (1-e)e & (1-e)^2 & (1-e)^2\\
e(1-e) & e(1-e) & e^2 & e^2\\
e^2 & e^2 & e(1-e) & e(1-e)
\end{array}
\right),
\label{eq:mat}
\end{eqnarray}
where its element $U_{ij}$ is the probability to observe state $i$ given the
previous state $j$. We calculate the stationary distribution over $n$ states
from the principal eigenvector $\vec{v}$ satisfying $U\vec{v} = \vec{v}$. For
notational convenience, we simply identify $\vec{v}$ with the stationary
distribution, assuming that it is normalized as a probability distribution.
Let $f_{ij}$ denote the inner product between $\vec{v}$ and
$\vec{m} \equiv (M_{CC}, M_{CD}, M_{DC}, M_{DD})$ with $i$=AllC and $j$=TFT.
The Perron-Frobenius theorem tells us that
$\overline{F_{ij}}(T)$ converges to $f_{ij}$ as $T \rightarrow \infty$.
Generalization to other strategies is straightforward.

We require that a successful strategy $k$ should satisfy the
following criteria:
\begin{enumerate}
\item Efficiency: It must achieve mutual cooperation if the co-player employs
the same strategy. It means that
$f_{kk}$ should approach $M_{CC}$ as $e \rightarrow 0$.
\item Distinguishability: It must be able to exploit AllC to avoid the neutral
drift. To be more precise, when $j$=AllC, the payoff difference
$f_{kj}-f_{jk}$ should remain positive finite as $e \rightarrow 0$.
\item Defensibility: It must not be exploited repeatedly by the co-player.
In other words, for strategy $k$ to be defensible, it should satisfy the
following inequality:
\[
\sum_{\tau=0}^{\nu-1} \left[ F_{kj}^{(t+\tau)} -
F_{jk}^{(t+\tau)} \right] \ge 0
\]
against any finite-memory strategy $j$, where the payoffs are
evaluated along every possible loop $S_t \rightarrow S_{t+1} \rightarrow
\ldots \rightarrow S_{t+\nu}$ allowed by $k$ and $j$.
\end{enumerate}

Let us explain those criteria in more detail.
The first and second criteria are relatively easy to check,
because
we only have to match each of the $N=2^{2^{2m}}$ strategies against itself and
AllC, respectively. For each of such pairs, we carry out the linear-algebraic
calculation of the principal eigenvector as in Eq.~(\ref{eq:mat}) to obtain
the stationary distribution $\vec{v}$.
By looking at which states occupy the most probabilities as $e
\rightarrow 0$, we can tell whether each given strategy meets the
criteria.

As for the defensibility criterion,
let us begin by assuming that both Alice and Bob use deterministic
strategies with memory length $m$. As mentioned above, the players can be
trapped in a loop for a period of $O(1/e)\gg 1$.
If the loop gives a lower payoff to Alice than to Bob
in one cycle, we say that she is exposed to the risk of repeated exploitation.
The defensibility criterion means that the player's strategy should
not allow such a risky loop.
For example, TF2T is not defensible, because it can
be trapped in the following sequence of moves when it meets WSLS:
\[
\begin{array}{ll}
\mbox{TF2T} & CCDD~CCDD~\cdots \\
\mbox{WSLS} & DDDC~DDDC~\cdots,
\end{array}
\]
which is clearly disadvantageous to TF2T at every repetition.
In fact, the defensibility criterion does not restrict the
co-player's strategy to deterministic ones. Now suppose that Bob has adopted a
stochastic strategy with memory length $m$. For every possible loop with
nonzero probability, we can find a
deterministic memory-$m$ strategy for Bob to reproduce the loop. For this
reason, if a given strategy does not contain risky loops against any
deterministic memory-$m$ strategy, it is also unexploitable by stochastic
memory-$m$ strategies. Even if Bob uses a longer
memory than $m$, Alice can marginalize his strategy to reconstruct
an effective stochastic one with memory length $m$.
The conclusion is the following: If a given strategy has no
risky loops against an arbitrary deterministic memory-$m$ strategy, it is a
sufficient condition for defensibility.
Therefore,
for each strategy that pass the efficiency and distinguishability criteria,
we examine all its loops by matching it against $N=2^{2^{2m}}$ deterministic
strategies one by one with probing every possible initial state.
The computational cost is bigger than those of
the other two criteria. Still, it is far less than $N^2$ needed
for direct
pairwise comparison, because only a small number of strategies pass
the efficiency and distinguishability criteria.
In practice, it is convenient to break up the defensibility test into two
parts: We first check if a given strategy leads to mutual defection against
AllD. If it does, we then proceed to check all its loops.

It turns out that no strategy with $m \leq 1$ satisfies these criteria
together. We thus proceed to $m=2$ to consider $N=2^{16}$ strategies. Then, our
numerical calculation shows that the
joint application of all these criteria singles out only four
strategies,
which are identical except for moves at $(CD,CD)$ and $(DC,DC)$.

To resolve this four-fold degeneracy,
we have to consider a stronger form of the efficiency criterion:
For this criterion
to hold true in general, mutual cooperation should be restored even if
the players, adopting the same successful strategy, go from $(CC,CC)$ to
$(CD,CD)$ via \emph{simultaneous} mistakes with probability $e^2$.
This argument determines the moves at $(CD,CD)$ and $(DC,DC)$ as follows:
Suppose that both the players choose $D$ at $S_t = (CD,CD)$. It results in
mutual defection $(DD,DD)$
and one must keep defecting at $(DD,DD)$ to be defensible against AllD.
In short, defection at $(CD,CD)$ is not the correct choice to recover
cooperation.
It is therefore clear that the successful strategy must prescribe $C$ at $S_t =
(CD,CD)$, which leads to $S_{t+1} = (DC,DC)$. Once again, unless the players
choose $C$ here, they will take the following undesired path:
\[
(CC,CC) \rightarrow [(CD,CD) \rightarrow (DC,DC)] \rightarrow [(CD,CD)
\rightarrow (DC,DC)] \rightarrow \ldots,
\]
where the square brackets denote a repeating loop of length two.
To sum up, $C$ is the correct choice both at $(CD,CD)$ and $(DC,DC)$, and the
recovery path from simultaneous mistakes goes as follows:
\[
(CC,CC) \rightarrow (CD,CD) \rightarrow (DC,DC) \rightarrow (CC,CC).
\]
Now, we have fully determined the move at every possible state.
The resulting strategy is tabulated in Table~\ref{table:str2}.
Surprisingly, it can be understood as a simple combination of TFT and ATFT.
Let us therefore call it TFT-ATFT and explain the reason in the next section.

\begin{table}[!ht]
\caption{List of moves in TFT-ATFT. This table
shows the proposed moves at time $t$ when the state is given as
$(A_{t-2} A_{t-1}, B_{t-2} B_{t-1})$, where $A_{t-2}$ and
$A_{t-1}$ ($B_{t-2}$ and $B_{t-1}$) are the focal player's
(the other player's) moves at the last two steps, respectively.
The underlined moves are the same as prescribed by TFT.
The states with the dagger symbol are related to
the two players' simultaneous mistakes (see text).}
\begin{tabular*}{\hsize}{@{\extracolsep{\fill}}cc@{\hspace{1cm}}cc}
\hline\hline
state & move & state & move
\\\hline
($CC$,$CC$)&\underline{$C$} &($DC$,$CC$)&\underline{$C$}\\
($CC$,$CD$)&\underline{$D$} &($DC$,$CD$)&\underline{$D$}\\
($CC$,$DC$)&$C$             &($DC$,$DC$)$^\dagger$&$C$\\
($CC$,$DD$)&$D$             &($DC$,$DD$)&$C$\\
($CD$,$CC$)&$D$             &($DD$,$CC$)&$D$\\
($CD$,$CD$)$^\dagger$&$C$   &($DD$,$CD$)&$C$\\
($CD$,$DC$)&\underline{$C$} &($DD$,$DC$)&\underline{$C$}\\
($CD$,$DD$)&\underline{$D$} &($DD$,$DD$)&\underline{$D$}\\
\hline\hline
\end{tabular*}
\label{table:str2}
\end{table}

\section{How TFT-ATFT stabilizes cooperation}

\begin{figure}
\begin{center}
\centerline{\includegraphics[width=0.8\textwidth]{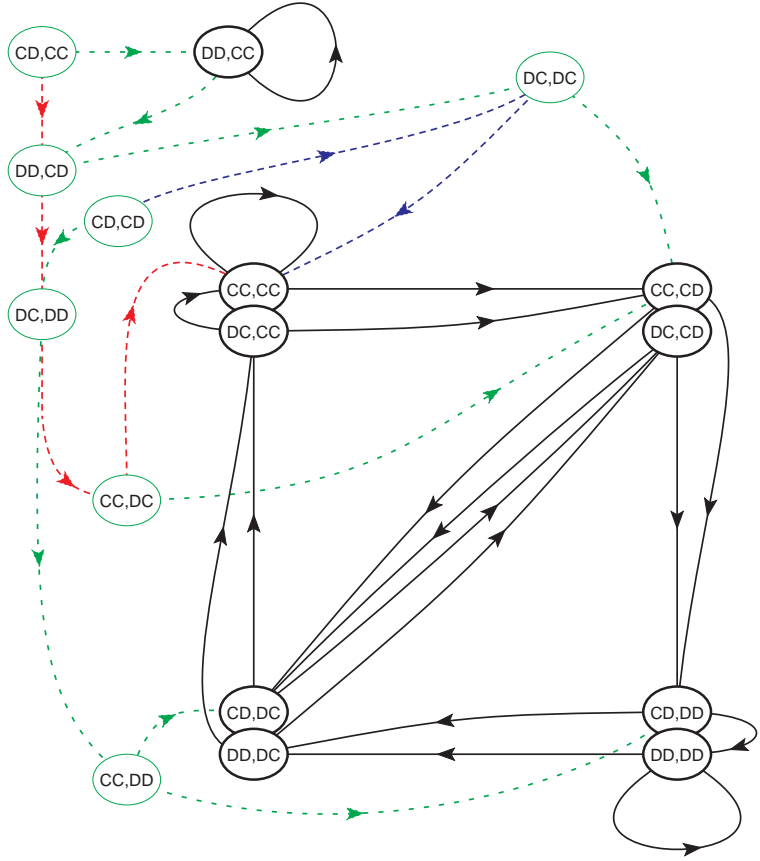}}
\caption{Visualization of the proposed strategy, TFT-ATFT. We represent
each state as a node and connect possible transitions between them.
For example, $(CD,CC)$ in the top-left corner is connected to
$(DD,CC)$ and $(DD,CD)$, because TFT-ATFT prescribes $D$ at $S_t = (CD,CC)$ so
that the next state should be either $S_{t+1}=(DD,CC)$ or $(DD,CD)$ depending on
the co-player's move.
The nodes drawn in thick black lines are recurrent and the others are transient.
The square cluster of eight recurrent nodes constitutes the pure
TFT to meet the defensibility criterion, and the other recurrent node at the
top, $(DD,CC)$, guarantees the distinguishability criterion by exploiting AllC.
The red dotted lines indicate how a player's implementation error is corrected
when both the players use this strategy. The blue dashed lines show another path
to recover cooperation when both defect by mistake.
The efficiency criterion is fulfilled by these two error-correcting
paths. Note that they make use of transient nodes, so that the co-player cannot
activate the error-correcting paths at will.
}\label{fig:itft}
\end{center}
\end{figure}

We have seen that a player's history of the IPD can
be understood as a series of transitions in a space of $n=2^{2m}$ states.
If Alice has memory length $m=2$, her strategy has prescribed her move $A_t$ at
state $S_t = (A_{t-2}, A_{t-1}, B_{t-2}, B_{t-1})$, by which it puts a
restriction on possibilities of the next state $S_{t+1} = (A_{t-1} A_{t},
B_{t-1} B_{t})$ as illustrated in Fig.~\ref{fig:transit}.
It is convenient to visualize a strategy as a transition graph~\cite{Nowak98}
by collecting all the possible transitions between states.
Suppose that Alice is using TFT-ATFT against Bob.
Figure~\ref{fig:itft} represents her strategy as a transition graph, in which
each node corresponds to a state $S_t$.
Recall that each state can be followed by two different states depending
on Bob's move $B_t \in \{C,D\}$. For this reason, every node in this graph has
outgoing links to two different nodes.
We stress that the resulting graph specifies all the possible transitions
allowed by Alice, regardless of Bob's strategy.

For any transition graph, we can classify the nodes into recurrent and transient
ones~\cite{Baek}: Bob can visit any of recurrent nodes repeatedly by choosing
$C$ and $D$ in an appropriate order. For example, $(CC,CD)$ is recurrent for
Alice's strategy in Table~\ref{table:str2} and Fig.~\ref{fig:itft}. It means
that Bob, starting from $(CC,CD)$, can return to it by choosing
$C$, $C$, and $D$ in sequence, generating the following path:
\[
(CC,CD) \rightarrow (CD,DC) \rightarrow (DC,CC) \rightarrow (CC,CD).
\]
On the other hand, transient nodes can be visited again only when Alice deviates
from her strategy. An example is $(CD,CC)$, which has no incoming links in
Fig.~\ref{fig:itft}.

Now, let us look into the three criteria from a different
perspective than in
the previous section. First of all,
if we wish to enforce a certain relationship between Alice's and Bob's payoffs,
as required by the defensibility criterion, we have to consider
zero-determinant
(ZD) strategies~\cite{Press,Hilbe12,Stewart,Traulsen,Szolnoki1,Szolnoki2}.
In the Appendix, we argue that TFT is indeed the only deterministic case of
the generous ZD strategies, which guarantees equal payoffs for both the players.
For this reason, the defensibility criterion essentially requires
that Alice
should normally use TFT, which does not allow repeated exploitation
no matter what Bob does.
Based on this observation, we enumerate accessible states from $(CC,CC)$ under
the assumption that Alice is using TFT:
\begin{equation}
(CC,CC) \rightarrow
  \left\{
  \begin{array}{l}
  (CC,CC)\\
  (CC,CD) \rightarrow
    \left\{
    \begin{array}{l}
    (CD,DC) \rightarrow
        \left\{
        \begin{array}{l}
        (DC,CC) \rightarrow \ldots\\
        (DC,CD) \rightarrow \ldots
        \end{array}
        \right.
    \\(CD,DD) \rightarrow
        \left\{
        \begin{array}{l}
        (DD,DC) \rightarrow \ldots\\
        (DD,DD) \rightarrow \ldots.
        \end{array}
        \right.
    \end{array}
    \right.
  \end{array}
  \right.
\label{eq:recur}
\end{equation}
One can readily check that this sequence is eventually closed with visiting
eight different states, which are underlined in Table~\ref{table:str2}. The use
of TFT as the default mode thereby determines moves at the eight different
states. All these states are recurrent.

However, pure TFT does not meet the efficiency criterion, and we
need
to accommodate it in the following way: If Alice made a mistake last time, she
should go to `Plan B' to correct it, which will work under the assumption
that Bob has
adopted the same strategy. At the same time, this process must be secured
against Bob's exploitation, because Bob may well become nasty to Alice. These
two requirements appear to be contradictory to each other.
Our point is that the security is ensured by making use of
transient nodes, which are out of Bob's control.
More specifically, the most probable scenario in Plan B
is that Alice visits $(CD,CC)$ by mistake with probability $e \ll 1$.
Let us suppose that Bob uses the same strategy and thus normally behaves as a
TFT strategist as shown above. Alice will choose $D$ at $(CD,CC)$ for the
following reason: If she did not, the next state would be $(DC,CD)$, one of the
recurrent states for which Alice's move is prescribed by the TFT
part [Eq.~(\ref{eq:recur})].
According to the prescription, however, they end up with
a series of TFT retaliation which Alice wants to avoid:
\[
(CD,CC) \rightarrow (DC,CD) \rightarrow (CD,DC) \rightarrow (DC,CD) \rightarrow
\ldots. \mbox{~(wrong)}.
\]
This is the reason that Alice's mistake must be followed by another $D$.
In general, we can construct a decision tree, which starts from $(CD,CC)$ and
branches off depending on Alice's choice, while Bob's moves are prescribed by
TFT. By pruning away `wrong' branches leading to inefficiency, we determine
Alice's correct moves. The result is as follows:
\begin{eqnarray}
&\underline{(CD,CC)}& \rightarrow (DC,CD) \rightarrow (CD,DC) \rightarrow \ldots
\mbox{~: TFT retaliation (wrong)}\nonumber\\
&\downarrow&\nonumber\\
&\underline{(DD,CD)}& \rightarrow (DD,DD) \rightarrow (DD,DD) \rightarrow \ldots
\mbox{~: mutual defection (wrong)}\nonumber\\
&\downarrow&\nonumber\\
&\underline{(DC,DD)}& \rightarrow (CD,DC) \rightarrow (DC,CD) \rightarrow \ldots
\mbox{~: TFT retaliation (wrong)}\nonumber\\
&\downarrow&\nonumber\\
&\underline{(CC,DC)}& \rightarrow (CD,CC) \mbox{~: back to the starting point
(wrong)}\nonumber\\
&\downarrow&\nonumber\\
&(CC,CC).&
\label{eq:scenario1}
\end{eqnarray}
The underlined states thus describe the correct path to recover mutual
cooperation in the first scenario of Plan B. It is identical to the red path in
Fig.~\ref{fig:itft}.

There also exists another
scenario that both Alice and Bob mistakenly choose $D$ with probability $e^2$.
The starting point will be $(CD,CD)$ this time, and the previous section
has already shown that mutual cooperation should be recovered int the
following way:
\begin{eqnarray}
&\underline{(CD,CD)}& \rightarrow (DD,DD) \rightarrow (DD,DD) \rightarrow \ldots
\mbox{~: mutual defection (wrong)}\nonumber\\
&\downarrow&\nonumber\\
&\underline{(DC,DC)}& \rightarrow (CD,CD) \mbox{~: back to the starting point
(wrong)}\nonumber\\
&\downarrow&\nonumber\\
&(CC,CC),&
\label{eq:scenario2}
\end{eqnarray}
provided that Alice and Bob have adopted the same strategy.
This corresponds to the blue path in Fig.~\ref{fig:itft}, and
completes the second scenario of Plan B.
In total, the efficiency criterion determines moves at six
different states,
underlined in Eqs.~(\ref{eq:scenario1}) and (\ref{eq:scenario2}).

We point out that Alice has chosen the \emph{opposite} to Bob's previous move
every time until reaching mutual cooperation at $(CC,DC)$ in
Eq.~(\ref{eq:scenario1}) or $(DC,DC)$ in Eq.~(\ref{eq:scenario2}).
In other words, the strategy means the following: Play TFT, but turn to ATFT if
you made a mistake last time, and return back to TFT when mutual cooperation is
recovered.
This interpretation is actually consistent with the use of $m=2$,
because your memory should be as long as two time steps at least,
to tell if you made a mistake last time. That is, to make a decision
between TFT and ATFT at time step $t$, you have to compare what you were
supposed to do as a TFT-strategist and what you actually did at
time $t-1$, the former of which is
encoded in the moves at $t-2$. Given the last time step only,
you would have no
way to judge whether your move was right.

Among the $n=16$ states, the remaining ones are $(DD,CC)$ and $(CC,DD)$. The
former state is accessed when Bob does not react to Alice's erroneous defection,
e.g., when Bob is using AllC. If that is the case, Alice as an ATFT strategist
will maintain $D$ at $(DD,CC)$ to gain a higher payoff than Bob on average:
\begin{equation}
(CD,CC) \rightarrow (DD,CC) \rightarrow (DD,CC) \rightarrow \ldots.
\label{eq:exploit}
\end{equation}
Due to this property, this strategy is able to exploit AllC, satisfying the
distinguishability criterion.
Now, the last state to consider is $(CC,DD)$. It is reached when Bob
defects at $(DC,DD)$. Then, Alice cannot maintain her
ATFT strategy, because it would make $(CC,DD)$ recurrent and thus violate the
defensibility criterion.
For this reason, Alice must return back to TFT at $(CC,DD)$ by choosing $D$:
\begin{eqnarray}
&(CC,DD)& \rightarrow (CC,DD) \rightarrow (CC,DD) \rightarrow \ldots
\mbox{~: indefensible (wrong)}\nonumber\\
&\downarrow&\nonumber\\
&(CD,DD).&
\label{eq:retaliate}
\end{eqnarray}
This completes the derivation of TFT-ATFT.

\section{Evolutionary dynamics}

Although we have been mainly concerned about the game between two players Alice
and Bob, let us consider an evolving population in this section.
The purpose is to see how our
designed TFT-ATFT performs in the presence of other strategies such as TFT and
WSLS in an evolutionary framework. Unfortunately, an investigation on the full
set of strategies with $m=2$ is unfeasible due to our limited computing
resources at the moment. As a preliminary test, this section checks the
performance of TFT-ATFT against the $16$ strategies with $m=1$.

\begin{figure}
\begin{center}
\centerline{\includegraphics[width=0.5\textwidth]{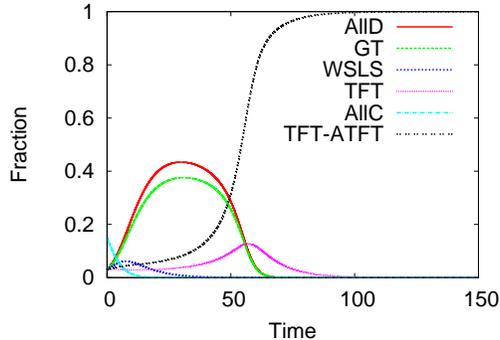}}
\caption{Numerical integration of the replicator equation. TFT-ATFT is included
in addition to the $16$ strategies with $m=1$. The payoff matrix $M$
[Eq.~(\ref{eq:payoff})] is
given by $M_{DC} = 1$, $M_{CC} = 1/2$, $M_{DD} = 0$, and $M_{CD}=-1/2$.
For every pair of strategies, we consider average payoffs over a long time
with error probability $e=10^{-2}$. The mutation rate is set as $\mu = 10^{-4}$.
}\label{fig:rd}
\end{center}
\end{figure}

By adding TFT-ATFT, we have $N=17$ strategies in total. The fraction of strategy
$i$ is denoted as $x_i$. The normalization condition gives $\sum_i x_i = 1$.
If TFT-ATFT is successful, its fraction will be of
$O(1)$ in the long run. In this evolutionary setting, our assumption is that
every player plays the IPD against each other for a long time to obtain an
error-average payoff.
To be more specific, we will construct the payoff matrix $M$
by choosing $M_{DC}=b=1$, $M_{CC}=b-c=1/2$, $M_{DD}=0$, and $M_{CD}=-c=-1/2$
for our calculation. The error probability is set to be $e=10^{-2}$ in
calculating error-averaged payoffs between every pair of
strategies. As in Sec.~\ref{sec:method},
let $f_{ij}$ denote the payoff that
strategy $i$ earns against strategy $j$ on average.
In a well-mixed population, the expected payoff of strategy $i$ is
calculated as $f_i = \sum_j f_{ij} x_j$. The replicator
equation~\cite{Taylor,Hofbauer,Smith} then describes
the time evolution of $x_i$ in the following way:
\begin{equation}
\dot{x}_i = (1-\mu)f_i x_i - \left< f \right> x_i + \frac{\mu}{N - 1} \sum_{j
\neq i} f_j x_j,
\label{eq:rd}
\end{equation}
where the left-hand side means the time derivative of $x_i$
and $\left< f \right> \equiv \sum_j f_j x_j = \sum_{ij} f_{ij} x_i x_j$ is the
mean payoff of the population. The last term on the right-hand side describes
mutation with rate $\mu$, which is set to be $10^{-4}$ in our numerical
calculation.
The replicator equation expresses an idea
that a successful strategy, relative to the population mean, will increase its
fraction in a well-mixed infinite population.
Equation~(\ref{eq:rd}) generates a trajectory on the $16$-dimensional space of
$(x_0, \ldots, x_{15})$, and $x_{16}$ is constrained by the normalization
condition. Under the deterministic dynamics of Eq.~(\ref{eq:rd}), the trajectory
generally depends on the initial condition.
For this reason, we have tested various initial conditions by taking
$4,854$ grid points with $x_i = 1/33, 5/33, \ldots, 17/33$ in
this $16$-dimensional space. Our numerical calculation shows that TFT-ATFT
occupies virtually the entire population for $99.3\%$ of the initial conditions.
Figure~\ref{fig:rd} depicts one such example.

Our simulation has shown that TFT-ATFT performs well even when many
strategies compete simultaneously. One might say that the success of TFT-ATFT is
not unexpected, because it has a longer memory than anybody else.
However, a longer memory does not necessarily lead to better performance.
The polymorphic population of memory-one strategies can be viewed as a
mixed-strategy player with $m=1$ from the viewpoint of TFT-ATFT, and
longer memory does not bring an advantage over a memory-one
stochastic strategy~\cite{Press}.
In this respect, it is not completely trivial that the
acquisition of additional memory bits makes such a huge impact on the
ecosystem of memory-one strategies.

\section{Discussion and Summary}

It is well known from the notion of direct reciprocity that cooperation can
constitute an equilibrium in the repeated prisoner's dilemma. However, when
players are in conflict with each other, as we are witnessing all around the
world, we have to instead ask ourselves how to lead them back to the cooperative
equilibrium. One may introduce population dynamics such as Eq.~(\ref{eq:rd}) to
achieve this goal. As members of a WSLS population,
for example, Alice and Bob can keep cooperating with each other, resisting
the temptation of defection with the aid of other members. However, we do not
know whether such a population always happens to exist around the players.
Moreover, even if it does exist, population dynamics can take place over a much
longer time scale than that of individual interactions.
In this work, we have tried to propose a solution such that recovers mutual
cooperation from implementation error within a time scale of individual
interactions.
The solution turns out to be a simple combination of TFT and ATFT. If Alice uses
this strategy against Bob, she normally uses TFT but shifts to ATFT upon
recognizing her own error.
Alice begins to behave as a TFT strategist again
when mutual cooperation is restored at $(CC,DC)$ or
$(DC,DC)$ [Eqs.~(\ref{eq:scenario1}) and (\ref{eq:scenario2})], or when Bob
keeps defecting after her cooperation [Eq.~(\ref{eq:retaliate})].
It is worth noting that Alice does not experience any security risk
by announcing what she plans to do. Predictability does not necessarily imply
exploitability. The point is that when designing a strategy, one should make use
of transient nodes, which the co-player cannot visit at will.

Knowing that Alice employs TFT-ATFT, Bob can also safely choose the same
strategy, considering that it is efficient and defensible. Having adopted
TFT-ATFT, he will find the following:
The efficiency criterion means that the average payoff $f_{ii}$
approaches
$M_{CC}$ as $e \rightarrow 0$ for $i=$TFT-ATFT. Due to the defensibility
criterion, furthermore, the average payoff $f_{ij}$
satisfies $f_{ij} \ge f_{ji}$ against an arbitrary strategy $j$. By definition
of the PD game, mutual cooperation is Pareto optimal, which means that $f_{ii}
\ge (f_{ij} + f_{ji})/2$. An immediate consequence is that $f_{ii} \ge (f_{ij} +
f_{ji}) /2 \ge (f_{ji} + f_{ji}) /2 = f_{ji}$.
For Bob, therefore, switching to another strategy $j \neq i$ does not bring him
any advantage. In other words, the strategy is the best response to itself and
thus constitutes a Nash equilibrium in the limit of $e \rightarrow 0$.

In addition, we remark the following three properties of TFT-ATFT:
First, the defensibility criterion holds true regardless of the
complexity of Bob's strategy, because this criterion solely
depends on
the transitions allowed by Alice (Fig.~\ref{fig:itft}).
Second, in terms of the stationary probability distribution $\vec{v}$,
the efficiency criterion implies
that $(DD,DD)$ has vanishingly small probability between two TFT-ATFT
players when $e \ll 1$. Indeed, players can escape from mutual defection
via erroneous cooperation (Fig.~\ref{fig:itft}):
\[
(DD,DD) \xrightarrow[\text{error}]{}
(DC,DD) \rightarrow (CC,DC) \rightarrow (CC,CC),
\]
with a time scale of $O(1/e)$.
Third, we may relax one of our assumptions that each player correctly perceives
her or his own payoff: Suppose that Alice sometimes miscounts her payoff,
by which she erroneously perceives Bob's cooperation as defection. However,
the discrepancy in their memories lasts only for $m$ time steps, after which
they remember the same history. They have a different state than
$(CC,CC)$ at this point,
but we have already seen that two TFT-ATFT players, from an arbitrary
initial condition, end up with mutual cooperation after $O(1/e)$
time steps. We thus conclude that cooperation based on TFT-ATFT is resilient to
perception error, if it occurs with a sufficiently longer time scale
than $O(1/e)$.

It is an interesting question if nature has already discovered TFT-ATFT, as has
been claimed in case of TFT~\cite{Wilkinson,Milinski,Stephens}. It is not
unreasonable, because this strategy has such a large basin of attraction under
the replicator dynamics as shown in the previous section. Even if that is not
the case, we believe that this strategy will find its own use in
applied game theory.
Of course, one should be careful at this point,
because our results have been obtained under highly ideal conditions. For
example, we have assumed that error occurs with probability $e \ll 1$ and that
two players interact sufficiently many times compared to $O(1/e)$, and any of
these assumptions can put serious limitation on the applicability of the
proposed strategy.

One may also think of designing more complex strategies with $m > 2$.
If we take, say, $m=3$, we have $n=64$ states and the number of possibilities
amounts to $N = 2^{64} \approx 2 \times 10^{19}$. It is beyond our
ability to check this larger set as we have
done in this work. Still, one may attempt to modify TFT-ATFT in several
directions by utilizing the extra bits of memory: For example, the recovery path
from erroneous defection may be shortened, and the possibility of
two successive errors may also be taken into account. Considering that TFT-ATFT
mainly refers to only the co-player's last move, however, we believe that the
modifications will have minor effects on the overall performance, as long as
they are based on the three criteria of efficiency, defensibility,
and distinguishability.

\section*{Acknowledgments}
We gratefully acknowledge discussions with Hyeong-Chai Jeong and Beom Jun Kim.
This work was supported by Basic Science Research Program through the
National Research Foundation of Korea (NRF) funded by the Ministry of Science,
ICT and Future Planning (NRF-2014R1A1A1003304).

\appendix
\section{TFT as a deterministic generous ZD strategy}

Following the common reformulation of the PD game in terms of
donation, let us parametrize the payoff matrix [Eq.~(\ref{eq:payoff})] by
setting $M_{DC} = b$, $M_{CC} = b-c$, $M_{DD} = 0$, and $M_{CD} = -c$, where $b$
and $c$ are the benefit and cost of cooperation, respectively, with $b>c>0$.
Then, a generous ZD strategy is represented by the following four
probabilities~\cite{Stewart}:
\begin{eqnarray*}
p_{CC} &=& 1\\
p_{CD} &=& 1 - \phi ( \chi b + c)\\
p_{DC} &=& \phi (b + \chi c)\\
p_{DD} &=& \phi (1-\chi) (b-c),
\end{eqnarray*}
where $p_{XY}$ denotes Alice's probability of cooperation when Alice and Bob did
$X$ and $Y$, respectively, at the last time step. The parameter $\chi$ must
satisfy $0 < \chi \le 1$ to produce a feasible generous
strategy. For this generous ZD strategy to be deterministic,
$p_{DD}$ must be either zero or one.
If $p_{DD} = 1$, we would get $p_{DC} = \frac{b+\chi c}{(1-\chi)(b-c)} >
\frac{b}{(1-\chi)(b-c)} > \frac{b}{b-c} > 1$, which does not make sense.
Therefore, we conclude that $p_{DD} = \phi (1-\chi) (b-c) = 0$, which implies
that $\phi = 0$ or $\chi = 1$. The former option should be discarded, however,
because it gives us a singular strategy $\mathbf{p} = (p_{CC}, p_{CD}, p_{DC},
p_{DD}) = (1,1,0,0)$, i.e., ``always cooperate or never
cooperate''~\cite{Press}. The other option, $\chi=1$, yields $\mathbf{p} =
(1,0,1,0)$, which is identical to TFT. Clearly, Alice and Bob then gain equal
payoffs~\cite{Stewart}.


\begin{thebibliography}{10}
\expandafter\ifx\csname url\endcsname\relax
  \def\url#1{\texttt{#1}}\fi
\expandafter\ifx\csname urlprefix\endcsname\relax\def\urlprefix{URL }\fi
\expandafter\ifx\csname href\endcsname\relax
  \def\href#1#2{#2} \def\path#1{#1}\fi

\bibitem{Sethi}
R.~Sethi, E.~Somanathan, Understanding reciprocity, J. Econ. Behav. Organ. 50
  (2003) 1.

\bibitem{Nowak}
M.~A. Nowak, Evolutionary Dynamics: Exploring the Equations of Life, Harvard
  University Press, Cambridge, 2006.

\bibitem{Szabo}
G.~Szab{\'o}, G.~Fath, Evolutionary games on graphs, Phys. Rep. 446 (2007)
  97--216.

\bibitem{Sigmund}
K.~Sigmund, The Calculus of Selfishness, Princeton University Press, Princeton,
  2010.

\bibitem{Perc}
M.~Perc, A.~Szolnoki, Coevolutionary games - a mini review, BioSystems 99
  (2010) 109--125.

\bibitem{Compare}
S.~K. Baek, H.-C. Jeong, C.~Hilbe, M.~A. Nowak, Comparing reactive and
  memory-one strategies of direct reciprocity, Sci. Rep. 6 (2016) 25676.

\bibitem{Axelrod}
R.~Axelrod, The Evolution of Cooperation, Basic Books, New York, 1984.

\bibitem{Molander}
P.~Molander, The optimal level of generosity in a selfish, uncertain
  environment, J. Conflict Resolut. 29 (1985) 611--618.

\bibitem{Boyd}
R.~Boyd, Mistakes allow evolutionary stability in the repeated prisoner's
  dilemma, J. Theor. Biol. 136 (1989) 47--56.

\bibitem{Nowak92}
M.~A. Nowak, K.~Sigmund, Tit for tat in heterogeneous populations, Nature
  (London) 355 (1992) 250--253.

\bibitem{Imhof05}
L.~A. Imhof, D.~Fudenberg, M.~A. Nowak, Evolutionary cycles of cooperation and
  defection, Proc. Natl. Acad. Sci. USA 102 (2005) 10797--10800.

\bibitem{Imhof07}
L.~A. Imhof, D.~Fudenberg, M.~A. Nowak, Tit-for-tat or win-stay, lose-shift?,
  J. Theor. Biol. 247 (2007) 574--580.

\bibitem{Imhof10}
L.~A. Imhof, M.~A. Nowak, Stochastic evolutionary dynamics of direct
  reciprocity, Proc. R. Soc. B 277 (2010) 463--468.

\bibitem{Dror}
Y.~Dror, Public Policymaking Reexamined, Transaction Publishers, New Brunswick,
  1983.

\bibitem{Kraines}
D.~Kraines, V.~Kraines, Pavlov and the prisoner's dilemma, Theory Decis. 26
  (1989) 47.

\bibitem{Nowak93}
M.~A. Nowak, K.~Sigmund, A strategy of win-stay, lose-shift that outperforms
  tit-for-tat in the prisoner's dilemma game, Nature (London) 364 (1993) 56.

\bibitem{Posch}
M.~Posch, Win-stay, lose-shift strategies for repeated games -- memory length,
  aspiration levels and noise, J. Theor. Biol. 198 (1999) 183--195.

\bibitem{Liu}
Y.~Liu, X.~Chen, L.~Zhang, L.~Wang, M.~Perc, Win-stay-lose-learn promotes
  cooperation in the spatial prisoner's dilemma game, PLoS ONE 7 (2012) e30689.

\bibitem{Young}
P.~Young, D.~Foster, Cooperation in the short and the long run, Game Econ.
  Behav. 3 (1991) 134--156.

\bibitem{Axelrod87}
R.~Axelrod, The evolution of strategies in the iterated prisoner's dilemma, in:
  L.~Davis (Ed.), Genetic Algorithms and Simulated Annealing, Morgan Kaufmann,
  Los Altos, CA, 1987, pp. 32--41.

\bibitem{Lindgren}
K.~Lindgren, M.~G. Nordahl, Evolutionary dynamics of spatial games, Physica D
  75 (1994) 292.

\bibitem{Sugden}
R.~Sugden, The Economics of Rights, Cooperation and Welfare, Blackwell, Oxford,
  1986.

\bibitem{Wu}
J.~Wu, R.~Axelrod, How to cope with noise in the iterated prisoner's dilemma,
  J. Conflict Resolut. 39 (1995) 183--189.

\bibitem{Boerlijst}
M.~C. Boerlijst, M.~A. Nowak, K.~Sigmund, The logic of contrition, J. Theor.
  Biol. 185 (1997) 281--293.

\bibitem{Hilbe09}
C.~Hilbe, Contrition does not ensure cooperation in the iterated prisoner's
  dilemma, Int. J. Bifurcat. Chaos 19 (2009) 3877--3885.

\bibitem{Nowak98}
M.~A. Nowak, K.~Sigmund, The dynamics of indirect reciprocity, J. Theor. Biol.
  194 (1998) 561--574.

\bibitem{Panchanathan}
K.~Panchanathan, R.~Boyd, A tale of two defectors: the importance of standing
  for evolution of indirect reciprocity, J. Theor. Biol. 224 (2003) 115--126.

\bibitem{Ohtsuki04}
H.~Ohtsuki, Y.~Iwasa, How should we define goodness? - reputation dynamics in
  indirect reciprocity, J. Theor. Biol. 231 (2004) 107–120.

\bibitem{Ohtsuki05}
H.~Ohtsuki, Y.~Iwasa, The leading eight: social norms that can maintain
  cooperation by indirect reciprocity, J. Theor. Biol. 239 (2006) 435–444.

\bibitem{Brandt}
H.~Brandt, K.~Sigmund, Indirect reciprocity, image scoring, and moral hazard,
  Proc. Natl. Acad. Sci. USA 102 (2005) 2666--2670.

\bibitem{Uchida}
S.~Uchida, Effect of private information on indirect reciprocity, Phys. Rev. E
  82 (2010) 036111.

\bibitem{Olejarz}
J.~Olejarz, W.~Ghang, M.~A. Nowak, Indirect reciprocity with optional
  interactions and private information, Games 6 (2015) 438--457.

\bibitem{Nowak90}
M.~A. Nowak, Stochastic strategies in the prisoner's dilemma, Theor. Popul.
  Biol. 38 (1990) 93--112.

\bibitem{Nowak95}
M.~A. Nowak, K.~Sigmund, E.~El-Sedy, Automata, repeated games and noise, J.
  Math. Biol. 33 (1995) 703–722.

\bibitem{Baek}
S.~K. Baek, B.~J. Kim, Intelligent tit-for-tat in memory-limited prisoner's
  dilemma game, Phys. Rev. E 78 (2008) 011125.

\bibitem{Press}
W.~H. Press, F.~J. Dyson, Iterated prisoner's dilemma contains strategies that
  dominate any evolutionary opponent, Proc. Natl. Acad. Sci. USA 109 (2012)
  10409--10413.

\bibitem{Hilbe12}
C.~Hilbe, M.~A. Nowak, K.~Sigmund, Evolution of extortion in iterated
  prisoner's dilemma games, Proc. Natl. Acad. Sci. USA 110 (2012) 6913--6918.

\bibitem{Stewart}
A.~J. Stewart, J.~B. Plotkin, From extortion to generosity, evolution in the
  iterated prisoner's dilemma, Proc. Natl. Acad. Sci. USA 110 (2013)
  15348--15353.

\bibitem{Traulsen}
C.~Hilbe, M.~A. Nowak, A.~Traulsen, Adaptive dynamics of extortion and
  compliance, PLoS ONE 8 (2013) e77886.

\bibitem{Szolnoki1}
A.~Szolnoki, M.~Perc, Defection and extortion as unexpected catalysts of
  unconditional cooperation in structured populations, Sci. Rep. 4 (2014) 5496.

\bibitem{Szolnoki2}
A.~Szolnoki, M.~Perc, Evolution of extortion in structured populations, Phys.
  Rev. E 89 (2014) 022804.

\bibitem{Taylor}
P.~D. Taylor, L.~B. Jonker, Evolutionarily stable strategy and game dynamics,
  Math. Biosci. 40 (1978) 145--156.

\bibitem{Hofbauer}
J.~Hofbauer, P.~Schuster, K.~Sigmund, A note on evolutionary stable strategies
  and game dynamics, J. Theor. Biol. 81 (1979) 609--612.

\bibitem{Smith}
J.~{Maynard Smith}, Evolution and the Theory of Games, Cambridge University
  Press, Cambridge, 1982.

\bibitem{Wilkinson}
G.~S. Wilkinson, Reciprocal food sharing in the vampire bat, Nature (London)
  308 (1984) 181--184.

\bibitem{Milinski}
M.~Milinski, Tit for tat in sticklebacks and the evolution of cooperation,
  Nature (London) 325 (1987) 433--435.

\bibitem{Stephens}
D.~W. Stephens, C.~M. McLinn, J.~R. Stevens, Discounting and reciprocity in an
  iterated prisoner's dilemma, Science 298 (2002) 2216--2218.

\end{thebibliography}

\end{document}